\documentclass{sig-alternate-05-2015}

\usepackage{etex} 

\usepackage[usenames,dvipsnames]{xcolor}
\usepackage{textcomp}
\usepackage{tikz}
\usetikzlibrary{shapes,arrows}
\usepackage{amsmath}
\usepackage{amssymb}
\usepackage{epsfig}
\usepackage[english]{babel}
\usepackage{url}
\usepackage{balance}
\usepackage{comment}
\usepackage{subfig}

\usepackage{enumitem}
\setlist[itemize]{noitemsep,
		  topsep=1pt,
		  parsep=0pt,
		  partopsep=0pt,
		  leftmargin=10pt,
		  }
\setlist[enumerate]{noitemsep,
		  topsep=1pt,
		  parsep=0pt,
		  partopsep=0pt,
		  leftmargin=10pt,
		  }

\usepackage{graphicx}
\usepackage{multirow}
\usepackage{tabu}
\usepackage{diagbox} 
\usepackage{pifont}
\usepackage{booktabs} 
\usepackage{multirow} 
\usepackage{graphicx} 

\newcommand{\tabred}{-2pt}

\newcommand*\rot{\rotatebox{90}}
\newcommand*\OK{\ding{51}}
\newcommand*\NO{\ding{55}}
\newcommand*\OKB{\ding{52}}
\newcommand*\NOB{\ding{56}}

\definecolor{lightblue}{rgb}{0.68, 0.85, 0.9}

\usepackage{color}
\usepackage{soul}
\newcommand{\hilight}[1]{#1}

\begin{document}
\title{DELTA: Data Extraction and Logging Tool for Android}

%

\numberofauthors{3} 
%
\author{
 \alignauthor 
 Mauro Conti\\
        \affaddr{University of Padua}\\
        \affaddr{Italy}\\
        \email{\normalsize{conti@math.unipd.it}}
 \alignauthor
 Elia Dal Santo\\
        \affaddr{University of Padua}\\
        \affaddr{Italy}\\
        \email{\normalsize{elia.dalsanto@studenti.unipd.it}}
 \alignauthor
 Riccardo Spolaor\\
        \affaddr{University of Padua}\\
        \affaddr{Italy}\\
       \email{\normalsize{rspolaor@math.unipd.it}}
}


\maketitle

\begin{abstract}
In the past few years, the use of smartphones has increased exponentially, and so have the capabilities of such devices. 
Together with an increase in raw processing power, modern smartphones are equipped with a wide variety of sensors and expose an extensive set of API (Accessible Programming Interface). 
\hilight{These capabilities allow us to extract a wide spectrum of data that ranges from information about the environment (e.g., position, orientation) to user habits (e.g., which apps she uses and when), as well as about the status of the operating system itself (e.g.,  memory, network adapters).}
This data can be extremely valuable in many research fields such as user authentication, intrusion detection and detection of information leaks.
For these reasons, researchers need to use a solid and reliable logging tool to collect data from mobile devices.

In this paper, we first survey the existing logging tools available on the Android platform, comparing the features offered by different tools and their impact on the system, and highlighting some of their shortcomings. 
Then, we present DELTA - Data Extraction and Logging Tool for Android, 
\hilight{which improves} the existing Android logging solutions in terms of flexibility, fine-grained tuning capabilities, extensibility, and available set of logging features. 
\hilight{We performed a full implementation of DELTA and we run a thorough evaluation on its performance. 
The results show that our tool 
has low impact on the performance of the system, on battery consumption, and on user experience.
Finally, we make the DELTA source code and toolset available to the research community.} 
\end{abstract}

\section{Introduction}
\label{introduction}

Thanks to their widespread adoption~\cite{SmartphoneOwnership2013} and extensive data collection capabilities~\cite{SurveyOnMobilePhoneSensing}, modern mobile devices, such as smartphones and tables, became very useful research tools \cite{SmartphonesToolsForSocialScientists, SmartPhoneSmartScience}. The flexibility and complexity of mobile Operating Systems are also increasing very quickly. 
From the simple embedded firmwares of early feature phones, modern mobile OSs now match the capabilities of traditional desktop systems. 
Mobile OSs such as Google's Android or Apple's iOS now support full-fledged multi-tasking, extensive APIs (Accessible Programming Interfaces), advanced connectivity, and a wide variety of hardware.

At the same time, mobile devices are equipped with a wide variety of sensors, \hilight{which collect data} from the surrounding environment.
\hilight{Built-in sensors such as gyroscopes, accelerometers, GPS receivers and digital compasses allow a mobile device to know its orientation, speed, and position. 
Similarly, light, pressure and temperature sensors can monitor the physical environment around the mobile device and apply on the fly corrections to system settings, such as adapting the display brightness to match the ambient one. 
As another example, built-in microphones and cameras collect audio and visual feedback.} 
Nowadays, all the aforementioned components (and more) are commonly included in modern mobile devices, usually even in low and mid-range ones.

The above mentioned hardware and software characteristics make modern mobile devices excellent data gathering devices for research purposes. 
In fact, they allow us to explore whole new areas of research, spanning multiple fields. 
For example, in the field of smartphone security, sensor and usage data allows the development of new authentication techniques~\cite{Touchalytics,MindHowYouAnswerMe,ISensedItWasYou}, user profiling studies~\cite{DetectingTargetedSmartphoneMalware}, and new attacks on user privacy that exploit side-channel information~\cite{AndroidSecuritySurvey,CantYouHearMeKnockingTIFS,Appscanner2016}. 
Another interesting avenue of research is using smartphones as portable monitoring stations, able to perform a variety of background monitoring tasks. 
Examples include collecting readings from personal health sensors~\cite{Ohmage}, recording ambient data like air and sound pressure, monitoring and tracing people movements and habits~\cite{Lifestreams}.
Software development can also benefit from such data, from simply collecting logs to monitor apps performance and crashes~\cite{AndroidCrashLogginFramework}, to analyzing user patterns and behavior~\cite{Lifestreams} in order to perform detailed usability tests.

We categorized the kind of data we can extract from mobile devices into three main categories:
\begin{itemize}
\item\textbf{Sensor data} is data we can gather directly by querying the many sensors embedded in modern mobile devices. 
This includes a wide variety of information about the device itself and its surrounding environment (e.g., the device's position, orientation and relative speed).
\item\textbf{Device/OS context data} is the state of the device itself and its operating system (e.g., the battery level, list of running processes, traffic statistics, and file system activity).
\item\textbf{User interaction data} is related to the device's user and her actions and habits, such as how she interacts with the touchscreen, with the keyboard and with elements of the User Interface.
\end{itemize}

\hilight{Given the high value of
this data for research}, a powerful and flexible multi-purpose logging tool would be of extreme value, as it would enables researchers to make data gathering for their projects \hilight{easier, effective and efficient}.



\paragraph{Contribution}
\label{our_contribution}

The contribution of this paper is twofold:
\begin{enumerate}
\item We present a survey on the existing logging tools for mobile devices. 
In particular, we select the more prominent tools present in the literature and we thoroughly compare their data collection features and architecture, highlighting both their selling points and limitations.
\item We  present DELTA - Data Extraction and Logging Tool for Android\footnote{DELTA is open source and it is available at \texttt{\url{https://github.com/tarockx/DELTA}}}, our own logging solution for the Android platform. 
We designed and implemented DELTA to cover the shortcomings most commonly found in other tools. 
Our tool logs as many information sources as possible, while at the same time allowing flexibility in what data is logged and at which frequency. 
Moreover, DELTA's architecture is designed to be modular and as non-invasive as possible in regards to user privacy and system security. 
\end{enumerate}

We make the DELTA source code and toolset available to the research community and practitioners, so that interested people can leverage it to streamline the process of logging data for their experiments.

\paragraph{Organization}
\label{organization}
The rest of this paper is organized as follows. 
In {Section~\ref{comparison_and_contribution}}, we present a thorough survey on the existing data collection tools for mobile devices. 
In Section~\ref{android_introduction} we introduce some basic concepts about the Android operating system and how its internal workings shaped the way we implemented our solution.
In Section~\ref{design_and_architecture} we go in-depth about the design and inner workings of the DELTA system, while in Section~\ref{evaluation} we sum up our results in terms of achieved goals and the security impact and energy requirements of our solution.
Finally, in Section~\ref{conclusions_and_future_work}, we draw conclusions and introduce some ideas for future works.

\section{Mobile Data Collection Tools}
\label{comparison_and_contribution}

In this section, we open with a survey on existing mobile logging tools and other related work in the literature (sections~\ref{existing_tools} and~\ref{additional_related_work}). 
We then sum up the main limitations common to existing tools (Section~\ref{limitations}). 
Finally, we explain how our proposal outperforms the existing solutions (Section~\ref{what_we_do_differently}).

\subsection{Existing tools}
\label{existing_tools}

While several logging tools exist that are aimed at a mainstream public (e.g., 
mSpy\footnote{mSpy - \url{https://www.mspy.com}}, 
KidLogger\footnote{KidLogger - \url{http://kidlogger.net}}, 
MobiStealth\footnote{MobiStealth - \url{http://www.mobistealth.com}}). 
but they are generally unsuitable for research purposes. 
Indeed, such tools are designed and marketed as ``spy apps" and they do not provide the precision and customization required for a research project. 
\hilight{In addition to that,} these apps are typically designed to send data to third-party remote servers, compromising the privacy of the user and blocking access to the collected data behind a pay-wall. 
For these reasons, we did not consider any ``spy apps" in this survey. 
Our solution is explicitly aimed at researchers, and we want it to be as open and easily accessible as possible.

Data collection is also an important aspect in the field of forensics. 
However, in forensic analysis the aim is to extract relevant information from a device at a certain point in time, usually in the context of a law enforcement operation. 
This is why forensic tools for mobile devices (e.g., ADEL~\cite{adel}) are designed to perform a one-time data extraction, rather than to continuously monitor a device.
This approach is very limiting for research purposes, where researchers typically want to collect usage data over time in order to find correlations and make predictions.
In fact, a one-time extraction cannot provide a history of sensor readings and system events because it is usually more focused at making a snapshot of the current state of a device.

In our comparison, we then focus on tools that can do continuous and/or periodic logging of data from more than a single source or sensor. 
Table~\ref{tab:comparison} summarizes a feature comparison between the tools we tested and our solution. We grouped the logging features in Table~\ref{tab:comparison} in a way that highlights the various contexts from which we can gather data from (e.g., sensor readings, screen interactions, network features). 
In the following paragraphs, we analyze each of these tools in detail.

\textbf{SystemSens} is an open source logging application presented by Falaki et al. in~\cite{SystemSens} that collects data from sensors and OS context, and it is designed to offer some extensibility options.
 However, this tool has some shortcomings that limit its usefulness for research purposes.
Firstly, it has a mainly monolithic design, with extensibility provided via AIDL\footnote{Android Interface Definition Language (AIDL) - \url{http://developer.android.com/guide/components/aidl.html}} (an Android inter-process communication mechanism) 
as an optional feature. Secondly, this tool uses a fixed global polling interval, set at a frequency of two minutes, which is often not sufficient for fine grained data analysis (more on these issues in Section~\ref{limitations}). 

Another similar tool is \textbf{DroidWatch}~\cite{DroidWatch}, an enterprise monitoring system for Android mobile devices. 
This tool concentrates on data sources 
such as phone call logs, visited websites, text messages and the user location. 
On the other hand, it ignores system information and sensor readings. 
This tool is also not natively extensible, and suffers from the same limitations as \textit{SystemSens}: no fine-tuning or advanced customization of logging behavior and a monolithic design.

\textbf{MobileSens}~\cite{MobileSens} is another app for logging user behavior in Android. 
The main aim of this tool is to profile user actions in order to study their behavior. 
Thus, logging is geared toward user actions and how they affect the state of the device (e.g., when the screen turns on/off, when the user sends messages). 
Like \textit{DroidWatch}, MobileSens is not extensible and no source code is provided. 

Developed by Wagner et al. at the University of Cambridge, \textbf{DeviceAnalyzer}~\cite{DeviceAnalyzer} is a comprehensive logging tool for Android. 
Among the Android tools we analyzed, DeviceAnalyzer is the one that logs the largest number of data sources. 
This tool is monolithic and not natively extensible, nor is it open-source, making it unsuitable for researchers that want to add their own custom logging plug-ins to the experiment. 
Secondly, \textit{DeviceAnalyzer} collects all data and merges it into a global data set, controlled by its authors. Although the authors provide access to the data set (on request), this makes it impossible for a researcher to deploy a specific, customized experiment to a specific group of users. 

\textbf{LiveLab}~\cite{LiveLab} is a tool similar to ours, built for the iOS operating system. 
This tool allows logging of various sensor readings and context data, with support for uploading it to a remote server. 
Similarly to other tools we examined, LiveLab does not provide fine-grained tuning of polling intervals. 
In addition to this, since apps not approved by the manufacturer cannot be installed on iOS devices, it requires an unlocked (``jailbroken'') copy of iOS. 
This requirement strongly limits the number of devices on which it can be deployed. 
For our implementation, we decided to target the Android operating system, to achieve maximum flexibility and guarantee a large potential user base. 
Contrarily to LiveLab, our tool is designed to require an unlocked (``rooted'') device only for certain advanced logging features, not obtainable through the standard operating system API.

\begin{table}[h!]
\centering
\scalebox{.8}{
\begin{tabular}{
	| >{\hspace{\tabred}}c<{\hspace{\tabred}} 
	| l<{\hspace{\tabred}} 
	| >{\hspace{\tabred}}c<{\hspace{\tabred}} 
	| >{\hspace{\tabred}}c<{\hspace{\tabred}} 
	| >{\hspace{\tabred}}c<{\hspace{\tabred}} 
	| >{\hspace{\tabred}}c<{\hspace{\tabred}} 
	| >{\hspace{\tabred}}c<{\hspace{\tabred}} 
	| >{\hspace{\tabred}}c<{\hspace{\tabred}}
	|}
\hline
\multicolumn{2}{|l|}{\diagbox{{\textbf{Features}}}{\textbf{Tool}}}  &  
\rot{SystemSens~\cite{SystemSens}}	& 
\rot{DroidWatch~\cite{DroidWatch}}	& 
\rot{MobileSens~\cite{MobileSens}}	& 
\rot{LiveLab (iOS)~\cite{LiveLab}} & 
\rot{DeviceAnalyzer~\cite{DeviceAnalyzer}} & 
\rot{\textbf{DELTA} }\\ 
	 
\hline
\multirow{9}{*}{\rot{Sensors}}	
&	Gravity sensor								& \NO	& \NO	& \NO	& \NO	& \OK	& \OKB\\
&	Accelerometer sensor						& \NO	& \NO	& \NO	& \OK	& \OK	& \OKB\\
&	Magnetic field sensor						& \NO	& \NO	& \NO	& \NO	& \OK	& \OKB\\
&	Proximity sensor							& \NO	& \NO	& \NO	& \NO	& \NO	& \OKB\\
&	Pressure sensor								& \NO	& \NO	& \NO	& \NO	& \NO	& \OKB\\
&	Light sensor								& \NO	& \NO	& \NO	& \NO	& \NO	& \OKB\\
&	Humidity sensor		 						& \NO	& \NO	& \NO	& \NO	& \NO	& \OKB\\
&	Log noise level	around device				& \NO	& \NO	& \NO	& \NO	& \NO	& \OKB\\
&	Record from microphone 						& \NO	& \NO	& \NO	& \NO	& \NO	& \OKB\\

\hline
\multirow{4}{*}{\rot{Screen}}	
&	Screen state (off/on/unlock)				& \NO	& \OK	& \OK	& \NO	& \OK	& \OKB\\
&	Touch events logging						& \NO	& \NO	& \NO	& \NO	& \NO	& \OKB\\
&	Keyboard state (open/close)					& \NO	& \NO	& \NO	& \NO	& \NO	& \OKB\\
&	Keylogging									& \NO	& \NO	& \NO	& \NO	& \NO	& \OKB\\

\hline
\multirow{10}{*}{\rot{System \& Power}}	
&	CPU statistics								& \OK	& \NO	& \NO	& \NO	& \NO	& \OKB\\
&	Battery statistics							& \OK 	& \NO	& \NO	& \OK	& \OK	& \OKB\\
&	Battery charging status						& \OK 	& \NO	& \OK	& \NO	& \OK	& \OKB\\
&	Memory statistics							& \OK 	& \NO	& \NO	& \NO	& \NO	& \OKB\\
&	System volume change						& \NO	& \NO	& \NO	& \NO	& \NO	& \OKB\\
&	Date / Time / Timezone changes				& \NO	& \NO	& \OK	& \NO	& \OK	& \OKB\\
&	Device turning on/off						& \NO	& \NO	& \OK	& \NO	& \OK	& \OKB\\
&	Storage space monitoring 					& \NO	& \NO	& \NO	& \NO	& \OK	& \OKB\\
&	File system activity monitoring				& \NO	& \NO	& \NO	& \NO	& \NO	& \OKB\\
&	Alarms ringing			 					& \NO	& \NO	& \NO	& \NO	& \OK	& \OKB\\
%
\hline
\multirow{4}{*}{\rot{Telephony}}	
&	Phone calls		 							& \OK	& \OK	& \OK	& \OK	& \OK	& \OKB\\
&	Incoming SMS messages						& \OK	& \OK	& \OK	& \OK	& \OK	& \OKB\\
&	Outgoing SMS messages						& \NO	& \OK	& \NO	& \OK	& \OK	& \OKB\\
&	Address book changes						& \NO	& \NO	& \OK	& \OK	& \NO	& \NOB\\
%
\hline
\multirow{13}{*}{\rot{Networking}}	
&	Airplane mode on/off	 					& \NO	& \NO	& \NO	& \NO	& \OK	& \OKB\\
&	Cell tower ID								& \OK	& \NO	& \NO	& \OK	& \OK	& \OKB\\
&	Cell signal strength	 					& \NO	& \NO	& \NO	& \NO	& \OK	& \OKB\\
&	WiFi connection info						& \OK	& \NO	& \OK	& \NO	& \OK	& \OKB\\
&	Scan of nearby WiFi HotSpots				& \OK	& \NO	& \NO	& \OK	& \NO	& \OKB\\
&	Network status (3g/WiFi/none)				& \OK	& \NO	& \NO	& \OK	& \OK	& \OKB\\
&	Network traffic statistics					& \OK	& \NO	& \OK	& \OK	& \OK	& \OKB\\
&	Network packet sniffing						& \NO	& \NO	& \NO	& \OK	& \NO	& \OKB\\
&	Opened URL logging							& \NO	& \OK	& \OK	& \OK	& \NO	& \NOB\\
&	Bluetooth state changes						& \NO	& \NO	& \NO	& \OK	& \OK	& \OKB\\
&	Bluetooth packet sniffing					& \NO	& \NO	& \NO	& \NO	& \NO	& \NOB\\
&	NFC device scanning							& \NO	& \NO	& \NO	& \NO	& \NO	& \NOB\\
&	NFC packet sniffing							& \NO	& \NO	& \NO	& \NO	& \NO	& \NOB\\
%
\hline
\multirow{6}{*}{\rot{Apps \& events}}	
&	Broadcast intents logging					& \NO	& \NO	& \NO	& \NO	& \NO	& \OKB\\
&	Running services							& \OK	& \NO	& \OK	& \OK	& \OK	& \OKB\\
&	Running applications and activities			& \OK	& \NO	& \OK	& \OK	& \OK	& \OKB\\
&	Foreground activity detection				& \NO	& \NO	& \NO	& \NO	& \NO	& \OKB\\
&	In-app UI interactions and changes			& \NO	& \OK	& \OK	& \NO	& \NO	& \OKB\\
&	App installs/uninstalls						& \NO 	& \OK 	& \OK	& \OK	& \OK	& \OKB\\
%
\hline
\multirow{3}{*}{\rot{Geoloc.}}	
&	Location services status					& \OK	& \OK	& \NO	& \NO	& \NO	& \OKB\\
&	Coarse location								& \OK	& \OK	& \OK	& \OK	& \OK	& \OKB\\
&	Precise location							& \NO	& \OK	& \OK	& \OK	& \NO	& \OKB\\
%
\hline
&	\textbf{TOTAL LOGGED FEATURES}				& 15	& 10	& 16	& 18	& 17	& \textbf{44}\\
%
\hline
\end{tabular}
}
\caption{Feature comparison table.}
\label{tab:comparison}
\end{table}

\newpage
\subsection{Additional related work}
\label{additional_related_work}
In this section, we discuss some additional tools and frameworks that perform logging operations on mobile devices. We did not include these solutions in our feature comparison, either because these solutions have a limited scope, or because they are not straightforward logging tools, but rather tools that log data to achieve a different primary goal.

One of the first works on device usage logging in the literature is \textbf{MyExperience}, targeted at the Windows Mobile platform and presented in~\cite{MyExperience}. It is a logging tool capable of collecting context data and performing actions in response to triggers configured by the experiment designer. Examples include presenting a survey to the user after a phone call has ended or when she is near a particular location. Even though targeted at an obsolete operating system, it is still relevant for historical reasons, being one of the first mobile device logging applications.

\textbf{DroidTracer}~\cite{DroidTracer} is a Linux kernel module that hooks into the Android system core at a low level. This tool aims to capture app interactions, logging data such as remote method calls between apps and Android API calls (e.g., access to the disk or the telephony services). In our tool, we followed the more traditional approach of leveraging (when possible) the Android API to collect our data. This gives us a wider array of logging possibilities, with the added advantage of not requiring root access.

An approach similar to \textit{MyExperience} is implemented by \textbf{Ohmage}, presented by Ramanathan et al. in~\cite{Ohmage}.  It is an Android tool designed to present the user with interactive surveys and self-monitoring tasks depending on triggers such as time and location. This tool is also capable of automatically collecting sensor data from the device, such as accelerometer, GPS, WiFi, microphone audio recordings and cell towers logs. This data is meant to give context in order to better understand the user's behavior, activities and surroundings, and is uploaded to a central server for analysis. We did not include it in our comparison due to its limited scope when it comes to the number and variety of logged data sources. Ohmage is also a tool designed with user interaction in mind, while ours is a passive tool that does not interfere with the user's activities.

An interesting and novel solution is presented by {Brouwers} et al. with \textbf{Pogo~\cite{Pogo}}, a middleware for mobile phone sensing. Its goal is to allow researchers to design sensing and data collection experiments and easily deploy them to Android phones. It provides a Javascript API, which exposes a limited subset of the Android API. This feature allows researchers to design experiments without being familiar with Java or the Android development platform. Pogo also sets an upper limit to the maximum resource usage for an experiment, to avoid depleting the battery too quickly. While the idea of using a simpler programming language to design experiments is an interesting one, it also makes Pogo very limited in its logging capabilities (which is the reason it is not included in our main survey). In fact, a lot of sensor and advanced contextual data can only be captured through calls to the native Android API.

\textbf{Dynamix~\cite{Dynamix}} is an open-source extensible context-sensing framework for Android. This tool is a plugin-based framework that collects sensor data, processes it to build a ``context", and then makes this context accessible to other applications via a dedicated API. While not strictly similar to our tool (it does not log data directly, but rather interprets and abstracts it to build user context information) its basic principles are similar to the ones present in DELTA. Dynamix can still be employed, in conjunction with a suitable app, to collect some user and system information at various levels of precision.

Finally, Cinque et al. in~\cite{AndroidCrashLogginFramework} propose a logging framework for the on-line failure analysis of Android smartphones. This consists of a framework that logs data related to app crashes and phone hangs/reboots. This data is then sent to a remote server to help in analyzing and troubleshooting bugs. Collected data includes the list of running applications and services, memory usage statistics, battery status and the current network connections. Such logging framework is not included in our main survey because it only logs a few predefined data sources, all related to app debugging.

\subsection{Limitations}
\label{limitations}
We encountered three limitations common to most existing tools. These limitations greatly reduce the usefulness of such tools in a research context. In fact, we found that researchers often opted for a custom solution, developed specifically for their project, instead of relying on existing logging tools~\cite{CantYouHearMeKnockingTIFS, ISensedItWasYou, MindHowYouAnswerMe}.

The first problem is that finding a tool that single-handedly satisfies all the data collection requirements for a particular research project is not easy. Most existing tools concentrate on one particular area (e.g., logging data from the device's sensors, collecting network packets). 
While a combination of different tools can be used to cover all the logging requirements of an experiment, this approach introduces some drawbacks: 
\begin{itemize}
\item \textit{Time consistency} - different tools will operate independently, without synchronization, and will timestamp data based on their internal timings. This can cause inconsistencies in the timestamps, which in turn can render the collected data difficult to correlate precisely and thus useless.
\item \textit{Sampling rate consistency} - different tools will most likely poll the sensors/APIs of the device at different frequencies. This causes potentially undesirable differences in the granularity of collected data.
\item \textit{Data format consistency} - different tools will use different formats to save the logged data. This means that a researcher would have to perform additional (and possibly non-trivial) post-processing on the data. This post-processing can cause inconsistencies when trying to correlate data together and is generally inefficient.
\item \textit{Data collection} - if different tools are used, there is no centralized mechanism to collect the gathered data and send it back to the researchers. This means that an additional ad-hoc implementation must be developed and deployed for this purpose.
\end{itemize}

The second common problem we encountered among existing tools is that they have limited support for fine tuning the sampling rates, in all those scenarios that require periodic polling of data (e.g., sensors reading).
Existing tools tend to set a predetermined (and often relatively long) polling interval.
While this limitation is usually implemented with the intent of reducing energy consumption, it can negatively impact on the usefulness of the collected data. 
In our approach, we want the \hilight{experiment designers} to be in control of the sampling rates of each logging operation, so that the usefulness of the collected data is maximized.

Finally, the tools we examined were not modular, instead using a monolithic design which did not provide an easy way to extend their logging capabilities. Even the ones that were extensible (e.g., SystemSens~\cite{SystemSens}) still had a monolithic ``core" that aggregated the basic logging features provided by the tool, with extensibility being an added extra. This approach leads to a lack of customizability and often violates the ``principle of least privilege", i.e., the app will often require more permissions than are actually needed to gather data for a certain experiment.


\subsection{What DELTA does differently}
\label{what_we_do_differently}
Given the above premises and limitations of the existing solutions, we decided to design a tool that would adhere to the following principles:
\begin{itemize}
\item \textit{Feature-richness} - our tool aims to provide a large variety of logging features out of the box, focusing on features that are not implemented (or not found together) in other logging tools.
\item \textit{Modularity} - we are aware that it would be impossible to provide support for every single logging need a researcher might have. Consequently, our tool is designed to be easy to extend through a dedicated plug-in system. This modular approach allows a developer to implement data collection features that are not available out of the box.
\item \textit{Fine-tuning} - our tool can be fine-tuned so that every single source of data can be polled at a configurable interval (or not polled at all). 
This approach allows the experiment designer to decide exactly which data sources she wants to monitor and at which frequency, which brings three advantages: (i) it maximizes the usefulness of the collected data; (ii) it reduces overhead by only tracking data that the experiment needs at the required intervals; (iii) the logger only requires the minimum privileges necessary to run each experiment.
\item \textit{Data distribution} - our tool can be configured to send the collected data back to a central server to make it accessible to the researchers that are running the experiment. This feature is important, as it lets the researchers access the data as it is collected, allowing them to monitor the logging process while it progresses. Automatic uploading also removes the need to physically retrieve the collected data from the devices (although this is still an option, which might be more suitable when dealing with large amounts of data).
\end{itemize}

Thus, we designed and implemented \textbf{DELTA - Data Extraction and Logging Tool for Android}, an extensible Android logging framework that aims to satisfy the aforementioned needs. 

\section{Android Operating System}
\label{android_introduction}

In this section, we briefly introduce some background knowledge and terminology about the Android Operating System. 
\hilight{In particular, we introduce some terminology and a few key concepts (that we refer to in the remaining part of the paper) about apps and the Android execution model, permission system and APIs.} 

\subsection{Android apps and execution model}
\label{android_applications_and_execution_model}

In the Android operating system, every application (``app" from now on) is distributed as a  \texttt{APK} file,
which contains all the app's components: bytecode, resources and additional metadata.
An \texttt{APK} file 
must be explicitly installed by the user, and the same goes when it comes to uninstalling an app.
When running, the code of an Android app is executed in a sandbox. 
This means that an Android app runs in isolation from the rest of the system, and cannot directly share memory with other apps. Any communication between different apps must be mediated by one of the inter-process communication techniques Android supports\footnote{Android application security - \url{https://source.android.com/security/overview/app-security.html}}. 




\subsection{System APIs and Permission System}
\label{system_apis_and_permissions}
\hilight{The Android OS provides an extensive set of APIs (Accessible Programming Interfaces), which mediate access to system resources 
and services. } 
While some APIs can be invoked freely, the ones that give access to sensitive data or potentially privacy-violating services (e.g., camera, microphone) cannot.
Indeed, the access to these APIs is protected by the Android Permission System~\cite{AndroidPermissionsDocumentation,AndroidPermissionsDemystified}. 
In practice, the operating system defines a series of permissions 
that an app must obtain in order to access some of the more sensitive APIs. 
For an app to be granted a certain permission, the developer must declare the use of that permission in the app's Manifest, an \texttt{XML} file that contains metadata information about the app. 
The list of permissions that an app uses is shown to the user when installing the app, and cannot change while an app is installed on the device.


The permission mechanism is 
meant to reduce the damage that could be generated by a successful attack that manages to take control of an app, by limiting the resources that app's process has access to. 
For this reason, the Android developer guide always recommends adhering to the \textbf{principle of the least privilege}. 
Unfortunately, permission over-provisioning is a common malpractice, so much so that research efforts have been spent in trying to detect this problem~\cite{PermissionGap}. 
This problem is even more relevant in our context, as the logging apps we analyzed all tend to over-provision on permissions, even though a researcher might only need a fraction of the functionality offered by a logging tool.

\subsection{The ``plugin problem"}
\label{the_plugin_problem}
As stated in the Section~\ref{what_we_do_differently}, we wanted our system to be modular, so that researchers could easily implement additional logging capabilities besides the ones we provide out of the box. 
This also means that we wanted our logging app (from now on referred to as the \textit{logger}) to only include the bare minimum code to fulfill the needs of each experiment. 
In essence, this consists of using some technique that allows the logger to dynamically load additional code (i.e., a plugin) at runtime. 
A typical technique is using a class loader, like the one provided by the Java Virtual Machine \cite{DynamicClassLoadingInJVM}.
However, even though Android programming is done in Java, and the class loader is still usable to load code dynamically from an external file, the Android permission system can severely hinder its functionality.
In practice, even if the \textit{logger} can technically load additional code at runtime, this code will still run in the same process as the logger, and thus with its permission set. 
Consequently, if the \textit{logger} does not hold the required permissions to run the dynamically loaded code, it will not be able to execute it correctly.
This means that creating a plugin system for applications such as DELTA is not straightforward. 
Our tool is designed to log a wide spectrum of data, and access to this data is often mediated by a specific permission. 
There are two ways to overcome this problem at runtime, but they both come with some specific drawbacks:
\begin{itemize}
\item \textbf{Preemptive permission over-provisioning} consists in having the \textit{logger} greedily declare the use of all existing permissions and be able to execute any dynamically-loaded code that is permission-protected. 
The advantage is that this way, the logger will be able to load any plugin from an external file. 
However, this solution violates the principle of least privilege, which as we have seen can be a risk for security. 
Moreover, it is not an effective solution: it needs to be constantly updated to include new permissions and it does not cover other Manifest extensions like services or GUI components that a plugin might require.
\item \textbf{Full decoupling} consists of having the plugins required for an experiment installed as distinct, separate apps on the device, alongside the \textit{logger}. 
This way, the \textit{logger} can be a minimal skeleton that delegates the logging operations to the plugins. 
Then, the logger will communicate with them using one of Android's built-in inter-process communication mechanisms. 
This solution seems more reasonable than the former one, but it has some flaws of its own.
The first one is that plugins have to be installed on the device as separate apps, which can be extremely annoying to the user. 
Secondly, inter-process communication in Android relies on data serialization and message routing, which result in overhead and consequently in a higher impact on system performance and battery life.
\end{itemize}

\section{Our proposal: DELTA}
\label{design_and_architecture}
In this section, we present the system architecture of 
DELTA, and explain how our implementation deals with the plugin problem.
To better understand the specific needs and concerns of researchers and users, we set up two focus groups to help us define DELTA's architecture and user interface. The first group included researchers that had worked on projects that used smartphone data as input, as well as Android developers. This group focused on system architecture, and helped us develop a design that would be easy to use for researchers and easy to extend for developers. The second group focused on usability, and included people not necessarily familiar with Android, such as researchers, students and other department staff. Input from this group helped us with suggestions about the design and functionality of the smartphone-side user interface of DELTA.

From the input received from these two focus groups, we set the following goals that the DELTA system design should fulfill:
\begin{itemize}
\item \textit{Minimal knowledge required by experiment designers} - we believe our tool should be accessible to everybody, even if they are not knowledgeable about the Android platform. DELTA provides dedicated graphical tools and an automated build system, which make creating new experiments very straightforward.
\item \textit{Ease of extension for plugin developers} - DELTA uses a fully modular plugin-based architecture, allowing developers to easily extend its logging capabilities.
\item \textit{Simplicity and security for users} - DELTA was designed to be as easy-to-use and non-invasive of user privacy as possible, thus encouraging voluntary user participation.
\end{itemize}

\subsection{System model}
\label{system_model}
\begin{figure}[h]
\centering
\subfloat[Architecture of the DELTA system, highlighting its components and their interactions.\label{img:DELTA_Architecture}]{%
      \includegraphics[width=0.44\textwidth, trim={0mm 2.25cm 0mm 1.15cm},clip]{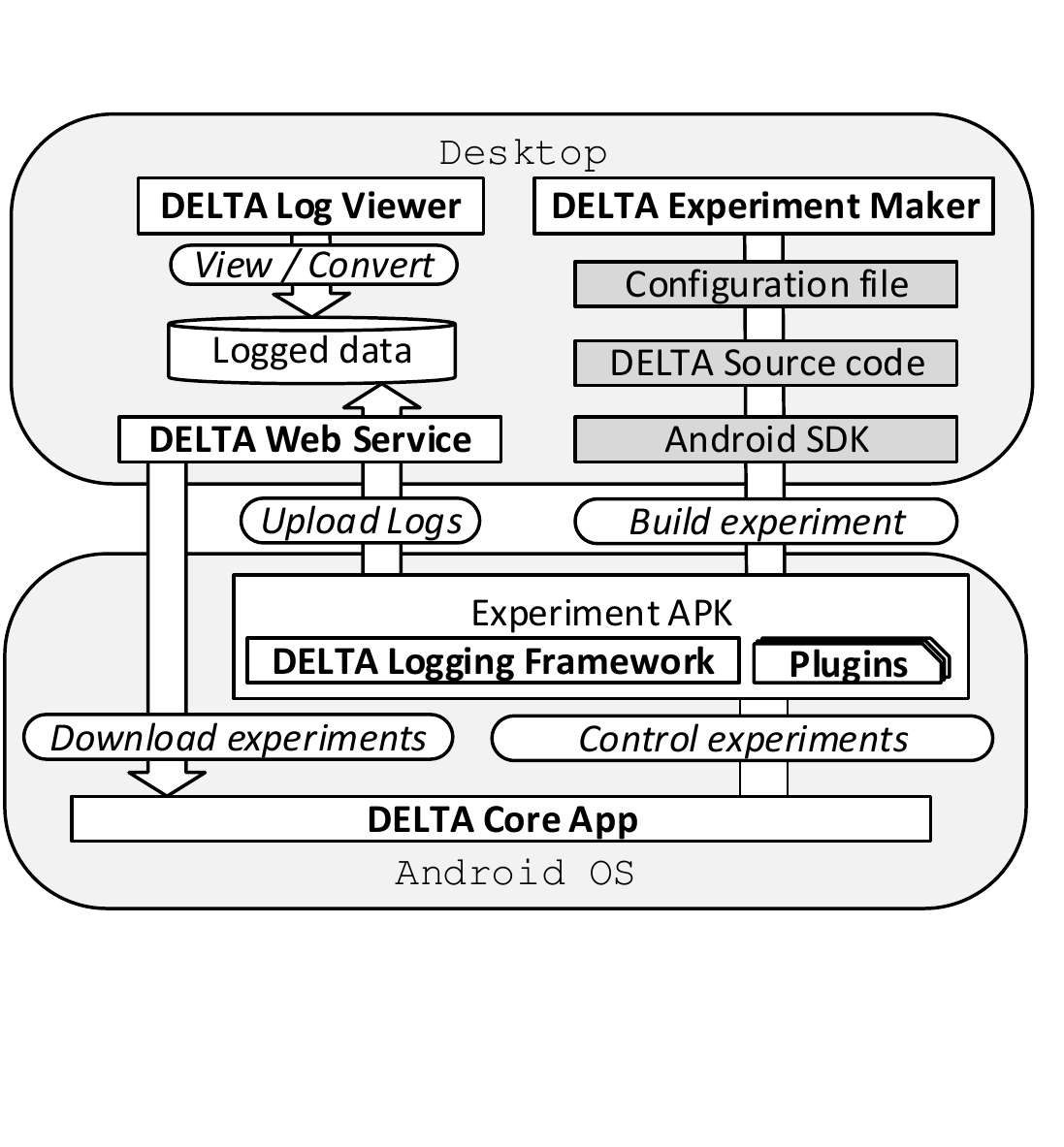}}
\\
\subfloat[Legend for all the system architecture diagrams.
	\label{img:Legend}]{\includegraphics[width=0.44\textwidth]{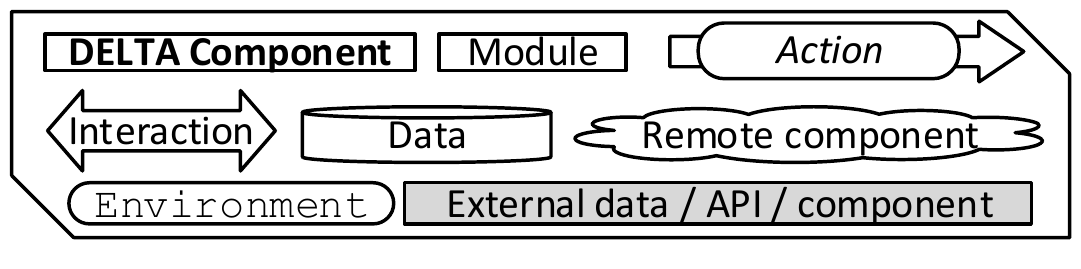}}
\caption{DELTA's architecture and legend.}
\label{img:whole1} 
\end{figure}

In this section, we illustrate the architecture of DELTA and show how each of the above mentioned goals (Section~\ref{design_and_architecture}) is achieved.
Figure~\ref{img:DELTA_Architecture} shows a high-level view of the DELTA system architecture, highlighting the various components. For a legend that explains the symbology of all diagrams in this section, refer to Figure\ref{img:Legend}.

The DELTA system allows users to configure, create, deploy and manage data logging experiments, in the form of standalone Android apps.
From now on we will refer to these apps as \textit{DELTA Experiments}, or simply \textit{experiments}. 
Each DELTA Experiment is a specialized instance of the DELTA Logging Framework component, plus a variable set of additional libraries, the DELTA plugins. Each plugin library contains one or more specialized modules that implement the actual data gathering operations. The DELTA Logging Framework is a generic framework that is able to instantiate and use the aforementioned plugins to schedule and perform the data logging and storage operations required by the author of the experiment. 

Researchers who want to create new experiments can do so with the DELTA Experiment Maker tool. The DELTA Experiment Maker is a desktop application that presents the users with a dedicated graphical interface, through which they are able to configure and build new DELTA Experiments. Depending on the chosen configuration, the DELTA Experiment Maker is able to build a custom DELTA Experiment that will include the minimal set of plugins to log the required data.  On the other hand, users that want to participate in experiments can use the DELTA Core App. The DELTA Core App is an Android app that is able to install, run, stop and remove DELTA Experiments on the user device. Optionally, the DELTA Core App also allows users to download new experiments from the web or send the logged data back to the researchers. These features are made possible by the DELTA Web Service, a simple self hosting web server that researchers can run to allow remote collection of the logged data and remote deployment of new experiments. Once the logged data has been collected, the DELTA Log Viewer desktop application can help researchers to merge it and convert it into different formats for better analysis.

\subsection{Logging Framework}
The DELTA Logging Framework is the core engine that runs DELTA's data gathering and storage operations. It is an Android app that implements a background service\footnote{Android services - \url{http://developer.android.com/guide/components/services.html}},  
the \textit{Logging Service}, which is in charge of running the experiment. When a researcher creates a new experiment a copy of the DELTA Logging Framework, together with a configuration file and all the necessary plugins, is compiled into a single \texttt{APK} file. This file is then deployed to the user's device, meaning the user only has to install a single package to run an experiment. This architecture allows DELTA to reap the benefits of a plugin system, while at the same time being as noninvasive as possible. In particular, only the strictly required plugins are included with the Logging Framework when building an experiment. Consequently, the generated \texttt{APK} will only require the bare minimum permissions. This minimizes overhead and complies with the principle of least privileges.
The anatomy and sub-components of a packaged DELTA Experiment are shown in Figure~\ref{img:Anatomy_of_a_DELTA_Experiment}.

%

The \textit{Logging Service} component is in charge of governing the experiment's life cycle. The framework itself does not implement any logging operations directly, delegating them to the various plugins instead.
The service autonomously maintains wakelocks and manages the timers that trigger periodic logging operations. Its \textit{Storage Manager} module implements facilities for timestamping, formatting and storing the logged data. This relieves plugin authors from having to manage such implementation details, so they only have to implement the routines that actually perform the logging of data.

Since the framework continuously runs in the background (if its experiment is running) it was important to minimize its impact on system resources, in order to save battery and not slow down the device. 
We designed the DELTA Logging Framework so it minimizes the time it keeps the device CPU awake. In particular, depending on the configuration of the plugins, we can have three cases:
\begin{itemize}
\item \textit{Plugins that do not require periodic polling}. These are plugins that log data reactively. In this case, no wakelocks are used.
\item \textit{Plugins that need to be polled periodically, but at long intervals}.
In case of plugins that do require polling, but are set to a polling frequency \textgreater10 seconds (the average time after which the CPU goes to sleep in Android devices), we use the energy-efficient alarms subsystem\footnote{Android alarms - \url{https://developer.android.com/training/scheduling/alarms.html}}. 
This facility allows us to schedule a periodic CPU wake-up without keeping the device constantly awake.  
\item \textit{Plugins that need to be polled at fast intervals}. In these cases, it is necessary to resort to wakelocks to keep the device awake, as the Alarms subsystem does not guarantee sufficient precision to schedule low-latency events.
\end{itemize}

The experiment creator can also decide to forego wakelocks altogether, and only let the experiment run when the device is awake. This is useful, for example, for experiments that only need to be running when the device screen is on.

Finally, the \textit{Storage Manager} module implements an internal cache, invisible to the plugins, in order to minimize I/O operations. On disk, data is stored using lossless \texttt{ZIP} compression, to reduce occupation of the user's storage. Once logged and stored, data can be either uploaded remotely, via the \textit{Data Upload} module, or dumped to a public portion of the user storage for easy manual extraction via the \textit{Data Dump} module. For more information on data storage and format, see Section~\ref{the_delta_logviewer}.

\subsection{Experiment Maker}
\begin{figure}[h!]
\centering
\subfloat[Components of a DELTA experiment package.
	\label{img:Anatomy_of_a_DELTA_Experiment}]{\includegraphics[width=0.4\textwidth]{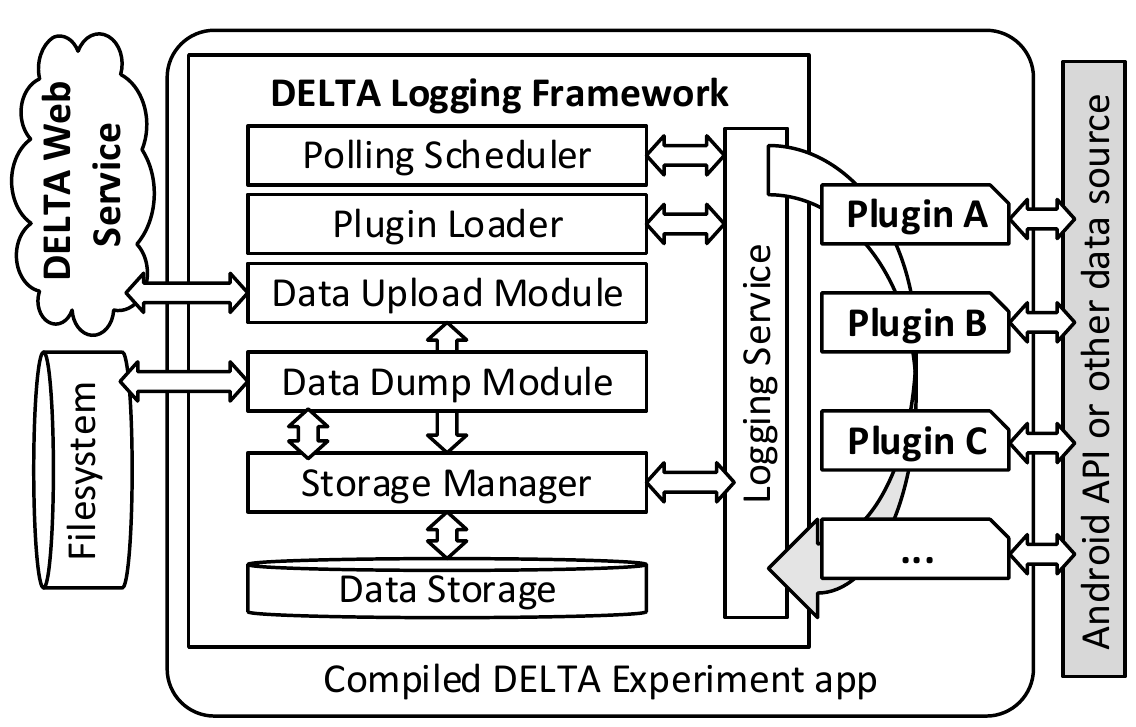}} 
\\
\subfloat[Architecture of the DELTA Experiment Maker and its main interactions.
	\label{img:ExperimentMaker_and_source}]{\includegraphics[width=0.4\textwidth]{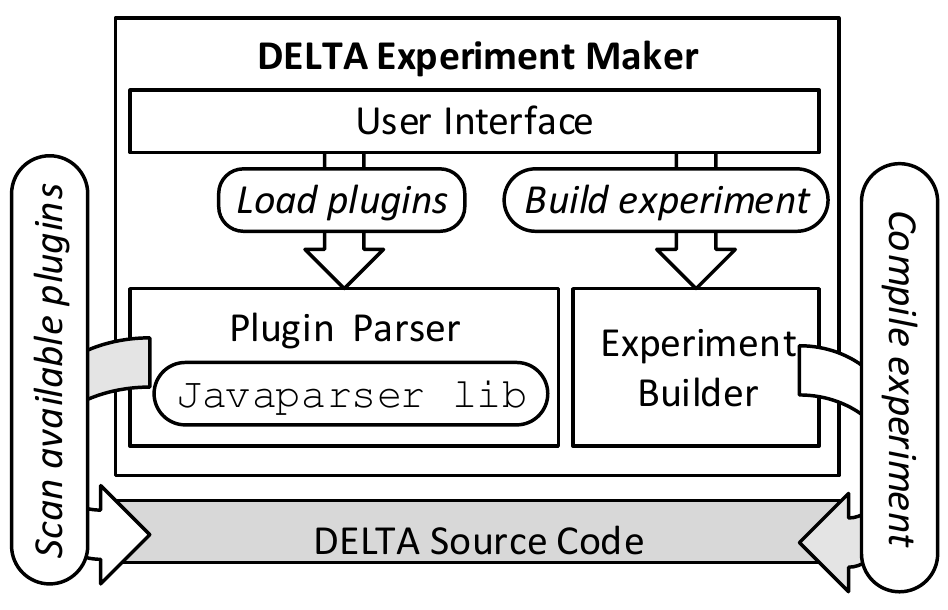}} 
\\
\subfloat[Architecture of the DELTA Core App and its modules.
	\label{img:DELTA_CoreApp_Architecture}]{\includegraphics[width=0.4\textwidth]{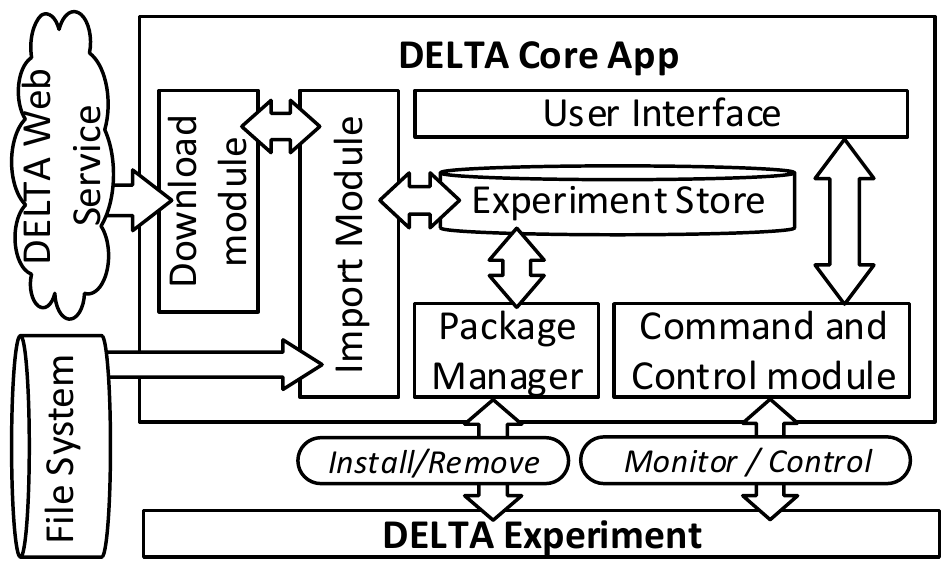}} 
\caption{DELTA's components.}
\label{img:whole2} 
\end{figure}

The main components of the DELTA Experiment Maker, and their interactions are shown in Figure~\ref{img:ExperimentMaker_and_source}. The DELTA Experiment Maker is a graphical cross-platform desktop tool, written in Java, that allows a researcher to easily create and configure a new DELTA Experiment. The Experiment Maker is designed so that the user can create a customized experiment with minimal instructions, without any knowledge of Android or Java programming.


The \textit{Plugin Parser} module leverages the \texttt{Javaparser} library to parse the DELTA source tree and detect all existing plugins. Using this information, the user interface presents the researcher with an automatically-populated list of available plugins. The researcher can then choose which plugins to include in the experiment, and is able to fine-tune the logging frequency of each one, plus any other advanced logging option. The Experiment Maker is also able to sign the experiments after building, using a user-provided certificate. This protects experiments against tampering.

Once the configuration has been decided, the \textit{Experiment Builder} module will invoke a series of custom build scripts to compile the experiment in a self-contained Android \texttt{APK} package. This package includes the main logging routines (the DELTA Logging Framework) and all (and only) the selected plugins.
%
%
%
%

%

\subsection{Core App}
The \textbf{DELTA Core App} is an Android app that manages DELTA Experiments, aimed at the end users that want to participate in experiments. Its architecture and main modules are shown in Figure~\ref{img:DELTA_CoreApp_Architecture}. 


The DELTA Core App provides an intuitive graphical interface, from which users can manage all the experiments available on their device. In particular, the app allows them to:
\begin{itemize}
\item Import new experiments, either directly from a file or by acquiring them through the DELTA Web Service (see Section~\ref{the_delta_webservice} for details).
\item Browse, install, remove and view details about stored experiments through the \textit{Experiment Browser} and \textit{Package Manager} modules. A detailed report is shown for each experiment, including information such as what the experiment logs, who is the author and the certificate used to sign it
\item Monitor the status of an experiment and send commands to it, using the \textit{Command and Control} module. These commands include the ability to start or stop the experiment at any time, plus additional commands to extract logged data or upload it to a remote server
\end{itemize}

The DELTA Core App maintains a list of running experiments, and is able to automatically restart them after the device is rebooted. This way, long-running experiments do not have to be manually restarted by the user in case they switch their phone off. The app is also able to automatically schedule periodic uploads of the logged data to a remote server, if the author of the experiment has enabled this functionality when configuring the experiment. To avoid depleting the user's mobile data plan, this function only runs if a WiFi connection is available.


\subsection{Web Service{}}
\label{the_delta_webservice}
The \textbf{DELTA Web Service} is a standalone self hosting web server, written in Java, that adds a couple of key features to the system:
\begin{itemize}
\item It allows users to \textbf{download DELTA Experiments} directly from within the DELTA Core App. This greatly facilitates the distribution of experiments, as researchers can deploy them without having to manually distribute the required files to all users.
\item It allows experiments to \textbf{send the logged data back to the researchers}. This means that researchers can collect the logged data automatically, without needing access to the user's device to perform a data dump. This also makes it possible for researchers to start analyzing incoming data while the experiment is still running, thus allowing them to speed up their analysis. It also has the added bonus of not clogging the user device with old data, since segments that are uploaded to the server are deleted from the local device cache.
\end{itemize}
The Web Service is provided alongside the main components, so any experiment author can run a copy of it independently. This ensures that the experiment author has total control over the collected data, which is not sent to a third-party server.\\
Note that using the Web Service is entirely optional. Experiment creators can still distribute experiment packages through any other mean (e.g., email, USB side-loading), the DELTA Core App is able to import them directly from the file system. Similarly, logged data does not have to be uploaded to the web service: it can simply be dumped to the device's public storage to then be extracted manually.


\subsection{Data format and the Log Viewer}
\label{the_delta_logviewer}

One of the advantages of having a multi-purpose logging tool like DELTA is that logged data has a consistent format, independent from the data source. In our implementation, we use \texttt{JSON} as our format for storing data. \texttt{JSON} is lightweight but at the same time, unlike the popular \texttt{CSV} format, it supports nested objects, so it can be used to represent more complex data.

For added flexibility, our data-reporting interface also supports logging of raw binary data. This comes in handy, as some plugins may log data that is not suited to be represented efficiently through strings (e.g., the audio recording plugin).

To avoid potential data corruption or loss, the DELTA Logging Framework stores data in separate chunks, which are then uploaded to the DELTA Web Service or otherwise retrieved by the interested party. To store data on disk, we use the \texttt{DEFLATE} algorithm. This gives good compression for textual data, without being too heavy on the device's CPU. This is important, since complex computations greatly affect the performance and battery life of mobile devices. Our tests have shown that the output from most of our plugins achieves an extremely high compression ratio, with compressed data typically ranging from 2\% to 6\% of the size of the original. All data is timestamped in milliseconds since Linux Epoch\footnote{00:00:00 UTC, 1 January 1970}.


The \textbf{DELTA Log Viewer} is a utility that can preview, merge and convert the logs collected by a DELTA Experiment. 
In particular, the user can choose to export just a specific data chunk, all the data logged from a specific device or all the data collected from all devices involved in the experiment. The DELTA Log Viewer can also convert data, if requested, from its native JSON format to a more user-readable CSV file format. DELTA Log Viewer can also be used to preview the collected logs, and to merge together any binary (non-textual) data collected during the logging process.

\subsection{Plugins and extensibility model}
\label{plugins_and_extensibility_model}
A \textbf{DELTA Plugin} is a specialized Java class that logs data on behalf of the DELTA Logging Framework. DELTA Plugins are contained in standard Android library packages (\texttt{AAR}), where each library package can contain one or more plugins. Contrary to what most other logging tools do, we designed DELTA to be completely modular, meaning that all the logging features we implemented are implemented as standard, optional DELTA plugins. No logging functionality is hardcoded in the DELTA Logging Framework which is, in fact, oblivious of the plugins. In our source code, we put plugins that require the same set of permissions together in the same library packages. This way we comply with the principle of the least privilege, as the final \texttt{APK} will only contain the strictly necessary libraries and thus will only require the minimum set of permissions to run the experiment.


From the point of view of a plugin developer, creating new plugins is a straightforward process, with an implementation footprint that consists of only a couple of standard interfaces and a Java annotation containing metadata (e.g., the author of the plugin, description of what it does, developer notes). This data is shown to experiment creators during the configuration phase. We also employ the Java annotation system to allow developers to easily define advanced options that the experiment creators can modify at configuration time, so that plugins can be flexible in how they log data.

There are two distinct kinds of DELTA plugins that can be created, depending on how the data is gathered. \textbf{Event plugins} log data reactively, i.e., they do not need to be polled periodically, and are useful to subscribe to system events or other data sources that can actively notify a plugin of content changes. \textbf{Polling plugins}, on the other hand, are polled periodically at a certain frequency, typically used to log data at a fixed rate. Examples include logging sensor readings or periodically collecting statistics about the system or the apps running on it.
\section{Evaluation}
\label{evaluation}
In this section, we begin by summing up the current state of the project in {Section~\ref{state_of_the_project}}. Then, we report how it impacts the user's devices in terms of security and privacy (Section~\ref{security_and_privacy}) and in terms of performance and energy consumption (Section~\ref{energy_and_performance}).

\subsection{State of the project}
\label{state_of_the_project}
At the time of writing, our implementation of the DELTA system is stable and fully functional. Table~\ref{tab:comparison} compares the list of logging features between DELTA and the other apps we analyzed. As reported in Table~\ref{tab:comparison}, DELTA is the most feature-rich of the analyzed apps, with 44 data sources logged. This is more than double the amount of data sources handled by the most feature-rich app (after DELTA).

DELTA also covers several advanced logging features that are rarely supported by other apps (or not at all). Examples include, among others, the logging of network packets, of the device's touchscreen, a full keylogger and logging of all available sensors in the Android framework. We consider this data among the most precious for scientific research, as it allows researchers to monitor network data and profile user habits, two features that are often essential in security-focused research.

It is worth noting that our implementation always strives to use official and documented APIs to collect data. This minimizes the risk of incompatibilities with future versions of Android and with devices we did not have a chance to test DELTA on. Our implementation is compatible with all Android 4.0.3 (and newer) devices, which covers more than 95\% of all Android devices in use today\footnote{Android dashboards - \url{https://developer.android.com/about/dashboards/index.html}}. 

Note that some of our plugins require root access (i.e, administration privileges) to work. These are plugins that log particularly sensitive data and have to rely on root privileges to bypass the standard Android API. In particular, affected plugins include the \textit{Touchscreen Logger}, the \textit{Packet Sniffer} and the \textit{Keyboard State logger}.

\subsection{Security and user privacy}
\label{security_and_privacy}
The very nature of DELTA raises security and privacy concerns for the user, as it is specifically designed to collect and store user data.
When developing DELTA, we considered several possible threats, and designed it with appropriate countermeasures:
\begin{itemize}
\item \textbf{User trust}. When installing an experiment on their device, users are installing an app that has the ability to log (potentially) sensitive data. It is crucial that the user trusts the author of the experiment. 
To achieve this, all experiment \texttt{APK}s are signed with the author's public key, and information on the corresponding certificate and certification authority is shown to the user before installing a new experiment.
The Android system itself will downright refuse to install the experiment package if the signature is missing or invalid.
\item \textbf{Data leak}. Considering the sensitive nature of collected data, it is important that this data is not easily accessible to external apps while still stored on the device. To achieve this, the DELTA Logging Framework will always keep logged data in its private storage space, where no other app can access it. Data is only dumped to the public storage space at a specific request from the user.
\item \textbf{Experiment hijacking}. The DELTA Core App, as we have seen, is used by the user to start, monitor and stop experiments. It is also used to send additional commands to the experiment, like a request to dump all logged data to disk or upload it to the DELTA Web Service. It is important that other apps are inhibited from sending such commands to the Logging Framework. To achieve this, we protected the framework service entry point with a custom signature-level permission\footnote{Android permissions usage - \url{http://developer.android.com/training/articles/security-tips.html#Permissions}}. 
This means that only the DELTA Core App can control experiments.
\item \textbf{Spoofing}. Lastly, we have to make sure that an external app cannot ``pass itself off" as a DELTA Experiment. In other words, we do not want an app that simply includes the DELTA Logging Framework to be displayed as an available experiment in the DELTA Core App. This could indeed confuse users, and potentially trick them into starting rogue experiments. This is why the Core App will only list experiments that the user has willingly imported from an external file or downloaded from the DELTA Web Service. There is no way to add an experiment ``silently" to the list of available ones
\end{itemize}

We feel that these measures should provide a reasonable level of security to the users of DELTA. 
We also want to point out that the security of DELTA can be compromised if the system itself is compromised, e.g., if a malicious app is running with root privileges. We feel, however, that this concern is irrelevant. If a malicious app is running on the device with root privileges, it can already collect user data directly and easily, without the need to extract it from the DELTA databases or hijacking the logging service itself.

\subsection{Energy and performance}
\label{energy_and_performance}
When it comes to performance, we designed DELTA to have a low overhead. The DELTA Logging Framework runs in a single process and, as we have seen previously, it employs various optimizations to reduce the load on the device's CPU and I/O interfaces.

Ultimately, it must be noted that the performance impact of DELTA is entirely dependent on the experiment configuration, and in particular on the number of plugins used and the polling frequencies they operate at.
A lot of the other logging tools we examined use a fixed (and typically very long) polling timer in an effort to save battery and CPU cycles. We wanted DELTA to offer researchers more flexibility when it comes to designing experiments, so we do not preemptively impose such restrictions. It is up to the researchers to balance the sample rates they need for their research against the impact such rates will have on the device's performance.

Another concern, always relevant in the field of mobile applications, is energy consumption.
We used a high precision power monitor to measure DELTA's impact on battery in a series of laboratory tests. Our test device was a Samsung Galaxy Nexus (GT-i9250) with only the stock set of apps installed (plus the DELTA Core App and various DELTA experiments). 

\begin{figure}[ht]
\centering
\subfloat[Display turned off.\label{img:power_off}]{
\includegraphics[height=2.8cm,width=.46\textwidth]{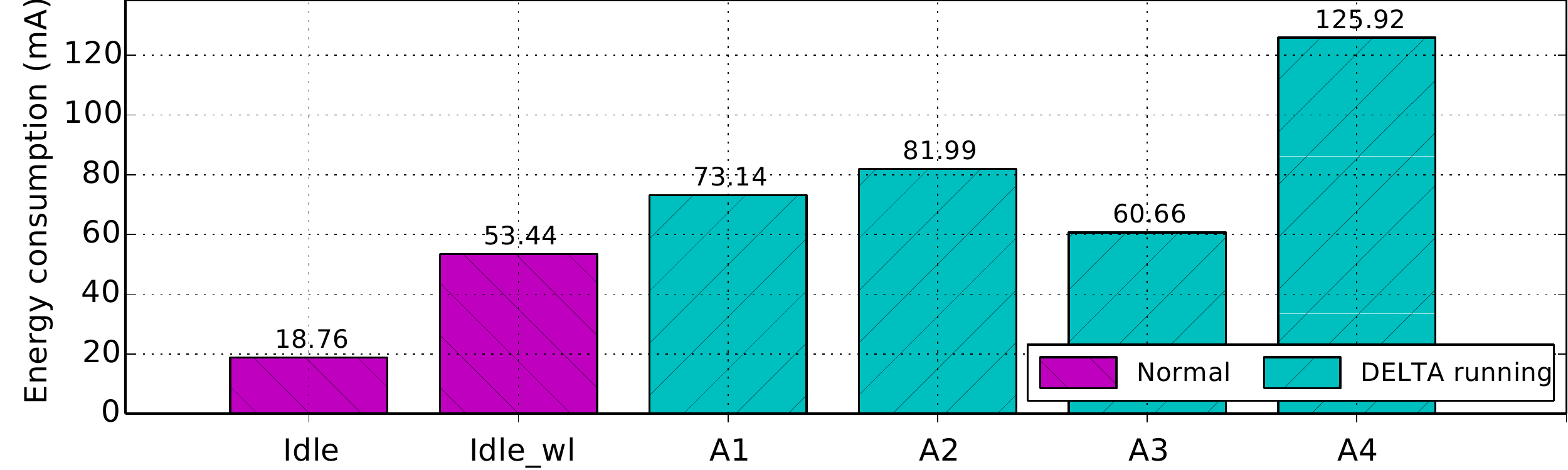}}
\\
\subfloat[Display turned on.\label{img:power_on}]{
\includegraphics[height=2.8cm,width=.46\textwidth]{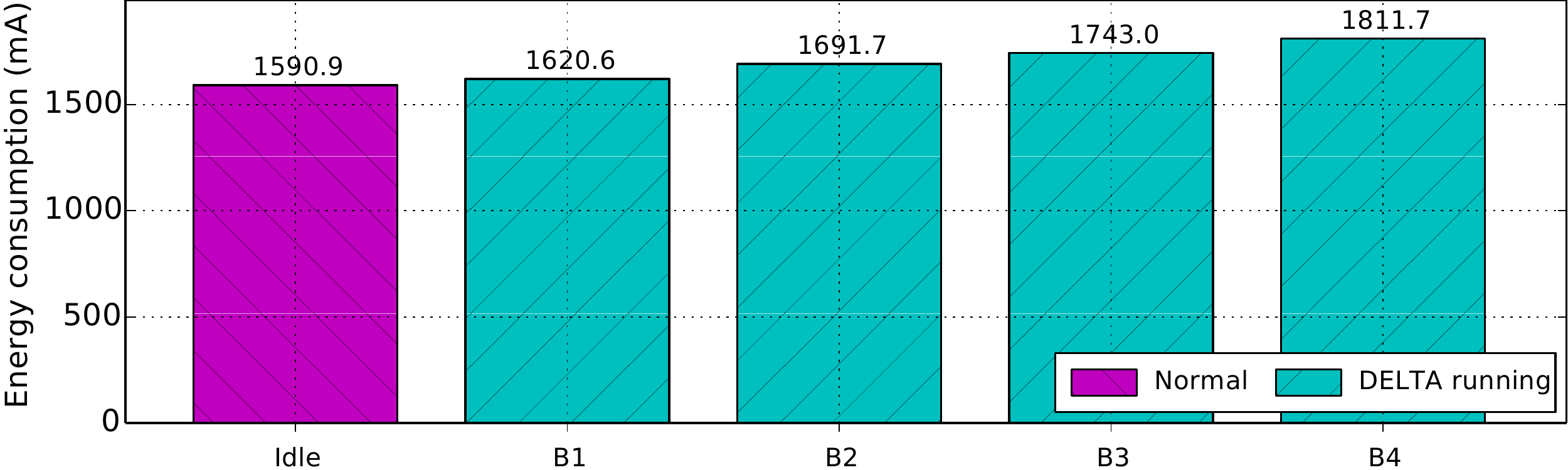}}
\caption{Average energy consumption with various DELTA experiments running.}
\label{img:power_drain}
\end{figure}

First, we ran a batch of tests with various experiments that executed while the device's screen was turned off, in order to assess the impact of DELTA on an otherwise idle device. Results are shown in Figure~\ref{img:power_off}, while Table~\ref{tab:experiments} explains the configuration of each of the tested experiments. We computed two separate averages while the device was not running experiments: one with the device completely idle and another with the device idle but wakelocked (i.e., prevented artificially from entering deep sleep). This is useful to understand how much of DELTA's energy drain comes from the device being awake and how much of it comes from the actual logging operations. From our preliminary tests, we noticed that the event plugins have a trivial impact on energy consumption. Thus, we concentrated on tests that queried polling plugins at frequent intervals and that kept the device constantly awake.
The results show some interesting findings:
\begin{itemize}
\item The device's wakeful state is usually the biggest factor that impacts energy consumption. Simply being awake adds about 35mA to the power footprint of the device.
\item Energy drain decreases rapidly as polling frequency increases. Experiment \texttt{A4}, which has a polling frequency of 50 milliseconds, drains the battery more than twice as fast as an idle wakelocked device. On the other hand experiment \texttt{A3} has a 200 milliseconds polling cycle, which results in a bare 12\% increase in energy drain. Decreasing the polling frequency further, makes the energy drain of a device running DELTA aligned with that of a simply wakelocked idle device.
\end{itemize}

The second batch of tests, whose results are shown in {Figure~\ref{img:power_on}}, was performed with the device in use. To get consistent results, we employed an automated script that simulated a user interacting with the device, by opening apps and simulating swipes and taps on the screen.
Since the device is constantly awake when the screen is on, we only have one reference reading, from the device running the script but not running any DELTA experiments (bar labeled as \texttt{Idle} in Figure~\ref{img:power_on}). For this batch of tests we chose plugins that would generate a lot of data and computational overhead, such as the touchscreen logger, the audio recorder and some sensor polling at high (50 milliseconds) frequencies. As can be seen, the various combinations of plugins generate an increase in energy drain ranging from 2\% to 13\% over the base value.

From the results of our tests, we can draw the following conclusions:
\begin{itemize}
\item Event plugins typically have a negligible impact on power consumption.
\item Keeping the device awake is one of the biggest factors in power drain. We want to underline that keeping the device awake is necessary to perform high precision polling, and thus the generated overhead would be present regardless of our implementation.
\item Energy consumption is tied directly to polling frequency. In particular, power drain spikes when polling occurs at a rate below 100 milliseconds, while it decreases rapidly when we poll less frequently. This is due to the fact that the Android system dynamically scales the frequency of the CPU, depending on the current computational load. Thus, faster polling means that the CPU does not have time to scale back down between one poll and the next, resulting in much higher energy drain.
\end{itemize}

\begin{table}[ht]
\centering
\scalebox{0.95}{
\small{
\begin{tabular}{
|>{\hspace{\tabred}}c<{\hspace{\tabred}}
|>{\hspace{\tabred}}c<{\hspace{\tabred}}
|m{.60\linewidth}|}
\hline
{\bf Label} & {\bf Min. poll} & \multicolumn{1}{c|}{{\bf Experiment}}                                                                  \\
\hline
\hline
\multicolumn{3}{| c |}{\textbf{Experiments with the screen turned off} (Figure~\ref{img:power_off})}                                                                                                  \\
\hline
Idle        & N/A                      & idle device, deep sleeping \\ \hline
Idle\_wl    & N/A                      & idle device, awake \\ \hline
A1          & 50 ms                   & one sensor polled \\ \hline 
A2          & 100 ms                   & two sensors polled\\ \hline 
A3          & 200 ms                   & one sensor polled \\ \hline 
A4          & 50 ms                    & network traffic polled\\ \hline 
\hline
\multicolumn{3}{| c |}{\textbf{Experiments with the screen turned on} (Figure~\ref{img:power_on})}                                                                                                   \\
\hline
Idle        & N/A                      & device in use (without DELTA)\\ \hline
B1          & N/A                      & touchscreen and keystrokes logged\\ \hline
B2          & 50 ms                    & B1 and two sensors polled\\ \hline 
B3          & N/A                      & B1 and audio recorded\\ \hline
B4          & 50 ms                    & B2 and audio recorded \\
\hline
\end{tabular}
}}
\caption{Overview of the experiment settings presented in Figure~\ref{img:power_drain}.}
\label{tab:experiments}
\end{table}

\section{Conclusions and future work}
\label{conclusions_and_future_work}
In this paper, we presented DELTA - Data Extraction and Logging Tool for Android, our implementation of a multi-purpose logging tool for Android. We started by comparing similar pre-existing tools, highlighting their target audience, features and common shortcomings.
Then, we showed how DELTA improves on existing solutions in terms of flexibility, customization, extensibility and logging scope. Of the solutions we examined, our solution is the only one that is built from the ground-up to be fully modular. DELTA is also the only one to achieve this without either violating the principle of least privilege or relying on inter-process communication. DELTA is also, by far, the most complete of the examined tools, logging more than forty different data sources.

Seeing as the energy problem is very important for mobile devices, we designed DELTA to have a low overhead, employing various techniques to minimize its energy impact. Our tests show how the DELTA Logging Framework is in itself lightweight, and how the majority of consumed energy either comes from the device's wakeful state or the data logging operations themselves.

We believe that DELTA's feature-richness and simple extensibility model can make it a precious tool for researchers. Writing a custom logging tool for an experiment is often a non-trivial endeavor, requiring time and knowledge about Android development. DELTA's feature-richness and high level of customization is often enough to create an experiment suited to a lot of data logging needs. When the basic set of plugins does not suffice, DELTA's modular design makes extension easy, abstracting away most of the complexities involved in developing a full-fledged custom logging tool.

Although our implementation is stable and fully working, there are ample opportunities for future extensions. One possible extension we consider very interesting is adding contextual awareness to DELTA. This idea consists of extending the DELTA Logging Framework's architecture so that plugins can be dynamically started and stopped, depending on context provided by other plugins. 
We would also like to expand the functionality of the DELTA Web Service, in order to allow researchers to configure and build an experiment directly from a web interface, rather than relying on a desktop application. 
This would allow us to provide DELTA as a service, instead of forcing researchers to install the DELTA Experiment Maker on their machines in order to create experiments.



\balance
\bibliographystyle{abbrv}
\bibliography{bibliography_slim}

\end{document}